# Liquid Crystalline Polymer Vesicles: Thermotropic Phases in Lyotropic Structures

Lin Jia[1#] and Min-Hui Li[1,2]*

[1]*Institut Curie-CNRS-Université Pierre et Marie Curie, Laboratoire Physico-Chimie Curie, UMR168, 26 rue d'Ulm, 75248 Paris, France.*

[2]*Institut de Recherche de Chimie Paris, UMR8247, CNRS - Chimie ParisTech (ENSCP), 11 rue Pierre et Marie Curie, 75231 Paris, France.*

\* Corresponding author: min-hui.li@chimie-paristech.fr

[#] Current address: *Department of Polymer Materials, Shanghai University Shangda Street 99, Mailbox 152, Shanghai 200444, China*

**Abstract**

This paper reviews the research work on the liquid crystalline (LC) polymer vesicles (polymersomes), where the thermotropic nematic and smectic phases are displayed in the lyotropic bilayer polymer membrane. LC polymersomes possess the properties of both liquid crystals and polymers, the two essential soft matters. LC polymersomes offer, on the one hand, novel examples of the interplay between orientational order and the curved geometry of a two dimensional membrane. Spherical, ellipsoidal and tetrahedral vesicles are discussed. On the other hand, LC polymersomes enable novel design of stimuli-responsive polymersomes using intramolecular conformational transition from nematic to isotropic phase of LC blocks. Photo-responsive polymersome bursting is highlighted.

Keywords: polymersomes; liquid crystal polymers; amphiphilic block copolymers; stimuli-responsive; nematic; smectic.

# 1. Introduction

Synthetic amphiphilic polymers have been largely developed since last decades for the purpose of forming self-assembled polymer vesicles (polymersomes) which mimick lipid vesicles (liposomes).

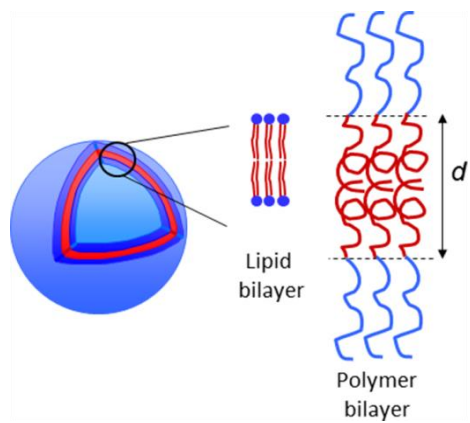

Figure 1. Illustration of a vesicle made of lipids (liposome) or amphiphilic polymers (polymersome). Because of the difference of molecular weights of the building blocks, the polymersome has a membrane thickness *d* superior to that of the liposome (*d* of liposome is of 3 – 5 nm, while *d* of polymersome is of 10 – 30 nm).

Polymersomes are much more stable, more robust and in most cases less permeable than liposomes due to the high molecular weight of polymers (Figure 1).(1) Another interesting feature of polymersomes is that their properties can be tuned extensively by chemical design of the amphiphilic building blocks.(2) (3) (4) (5) These nanostructures are currently studied as a means of drug delivery and biomedical imaging for their ability to entrap hydrophobic molecules in the membrane and encapsulate hydrophilic ones in the inner aqueous compartment.(6) The tailor-design of smart polymersomes, i.e., stimuli-responsive polymersomes bearing a protective coat, site-specific targeting ligands and a cell-penetrating function, is the state-of-the-art research in this field. (3) (7) (8) The research of our group has focussed, since several years, on the development of stealth and stimuli-responsive polymersomes made from amphiphilic block copolymers.(9) (10) (11) (12) (13) (14) We have developed responsive polymersomes by combining for the first time the properties of polymers with those of liquid crystals (LC) in the bilayer membrane. It is well known that liquid crystal systems excel as responsive systems and can respond to multiple stimuli including temperature, light, electric and magnetic fields. If this responsiveness could be retained in the liquid crystal membrane of polymersomes, these LC polymersomes would be stimuli-responsive smart polymersomes. For this purpose, thermotropic nematic and smectic LC polymers were used to constitute the hydrophobic core in order to trigger by physical stimuli the disassembling or the morphological change of the polymersomes. Poly(ethylene glycol) (PEG)-based polymer was chosen to construct the hydrophilic corona because of its biocompatibility and ability to reduce protein adsorption (stealthy towards immune systems). (15–18)

Amphiphilic copolymers in dilute solution of water (lyotropic system) can self-assemble into core-shell colloidal structures (generally called micelles). Typically, when the solvent becomes poorer for one of the polymer blocks, the formation of micelles will retrieve the insoluble block and hide it from the solvent to limit unfavorable interactions. Core segregation from aqueous milieu is the direct driving force for micellization and proceeds through a combination of intermolecular forces, including hydrophobic interaction, hydrogen bonding, electrostatic interaction and metal complexation of constituent block copolymers.(19) In this review, we will focus on the hydrophobic interaction, especially the anisotropic hydrophobic interaction in hydrophobic LC polymer blocks (and the additional orientational and positional orders resulted from it). Different morphologies such as spherical micelles, cylindrical micelles and vesicles are accessible by self-assembly of amphiphilic block copolymers in dilute solution. From the experimental point of view, the micelle morphology depends generally on the chemical structure of the copolymer, the hydrophilic/hydrophobic weight ratio, the solvent properties, the salt concentration, the solution pH and the temperature.(20–23) However, according to the nature of hydrophobic coral block, polymer micelles could be thermodynamically equilibrated or "frozen". In this last case, experimental conditions, such as the copolymer concentration, the rate that the solvent becomes poorer for one of the polymer blocks (e.g., the addition rate of a co-solvent) and the shear rate, also affect the final micelle morphology.(20–23) Take the hydrophilic/hydrophobic weight ratio as an example. Vesicles were formed in water by PEG-*b*-polybutadiene (PEG-*b*-PB or PEG-*b*-PBD) as its hydrophilic/hydrophobic weight ratio was around 35/65 (a phospholipid-like ratio),(1) while vesicles were also obtained by poly(acrylic acid)-b-polystyrene (PAA-*b*-PS) with a very short hydrophilic block (e.g. hydrophilic/hydrophobic weight ratio ~ 4/96) in a mixture of water and dioxane.(24) When a rod-like polymer block is introduced into a block copolymer, the shape anisotropy and additional order in the rod-like block (introduced by liquid crystalline or crystalline structures, or secondary structures such as α helices or β sheets in the case of a peptide) also influences the self-assembly.(25), (26), (27), (28), (29), (9), (30), (31), (32), (10), (33), (11), (34).

From the theoretical point of view,(35), (36), (37), (38) the morphology of an equilibrium micelle is governed by the free energy per chain (F) from three contributions: the F contribution of the core−corona interface, the F contribution due to the hydrophobic blocks in the core and the F contribution due to the hydrophilic blocks in the corona. In the case of crew-cut micelles made of flexible nonionic block copolymers, it has been shown that the elastic stretching of hydrophobic blocks in the core gives rise to micelle polymorphism (spherical micelles, cylindrical micelles, vesicles, etc.). Without this contribution, an equilibrium diblock copolymer micelle would always have a spherical shape. However, there are still a lot of unsolved problems even in the theory of nonionic polymer micelles.(38) For example, the effects related to different solubility of the components in solvent/co-solvent mixtures remain essentially unaddressed in the analytical theory of polymer micelles, while in experiment, mixtures of solvents are often used in micelle preparation protocol. Another example is about crystalline and liquid crystalline polymers. The theories of nonionic micelles focus mostly on the aggregates wherein core domain is in the amorphous state, while micelles with crystalline and liquid crystalline cores are rarely discussed.

LC polymer vesicles discussed in this review offer, on the one hand, novel examples of the interplay between orientational (and positional) order and the curved geometry of a two dimensional membrane, and on the other hand, novel design of stimuli-reposnsive polymersomes using intramolecular conformational transition from nematic to isotropic phase of LC blocks. The review consists of two parts. In the first one, we will discuss the different LC polymersomes studied by us, where LC polymer is nematic or smectic. The molecular organisation of mesogens and polymers in the polymersome membrane will also be detailed. In the second part, we will describe how to design a LC polymersome that can be opened suddenly by UV light using nematic LC polymer. We will show that the knowledge of the molecular organisation in the membrane is essential for this design. Thermo-sensitivity of LC polymersomes will also be discussed.

## 2. Liquid crystalline polymersomes

*2.1: LC amphiphilic block copolymers*

In the polymer vesicles developed by us, the hydrophobic block of the amphiphilic copolymer is not classical amorphous and flexible polymer, but a liquid crystalline polymer. Side-on and end-on side-chain LC polymers with different chemical structures have been synthesized (see Figure 2, Scheme 1 and Scheme 2).

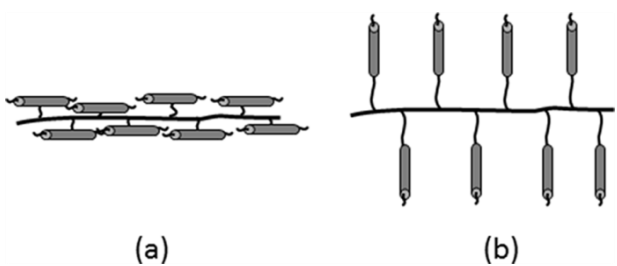

Figure 2. (a) side-on side-chain LC polymer and (b) end-on side-chain LC polymer. The small cylinders represent the rod-like mesogens (see Scheme 1 and Scheme 2).

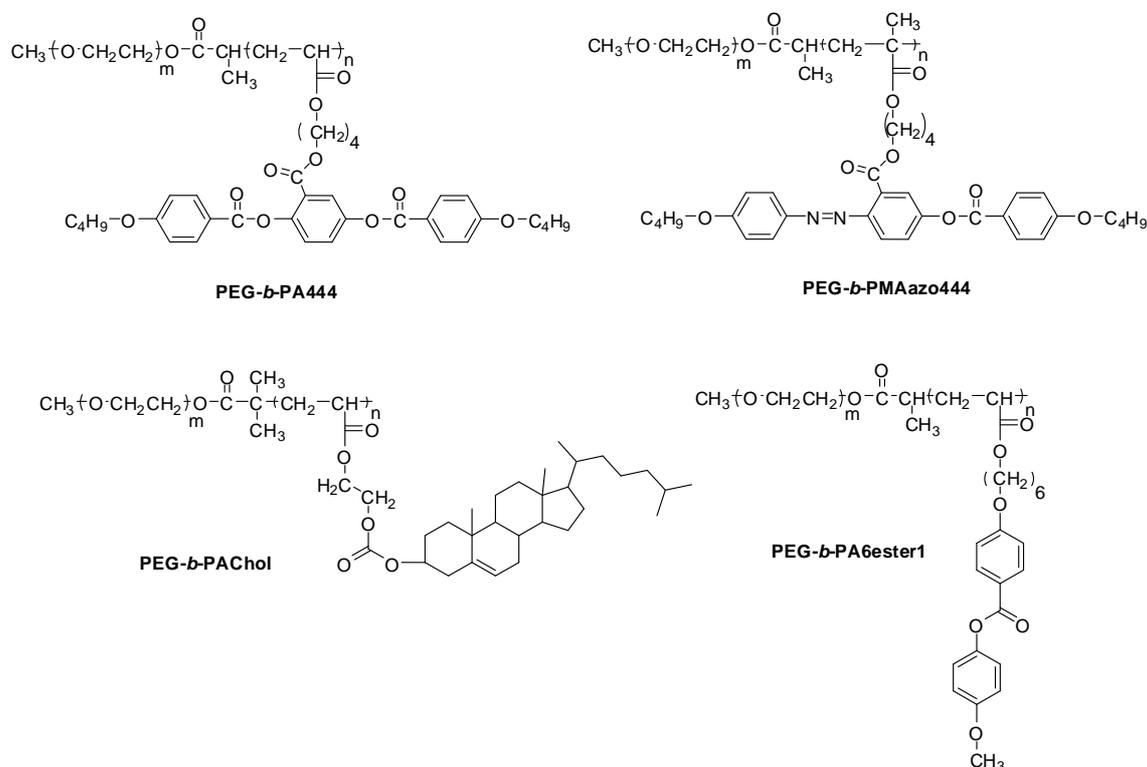

Scheme 1. Chemical structures of LC amphiphilic diblock copolymers synthesized by typical ATRP polymerization. The copolymers will also be noted as $P1_m$-$b$-$P2_n$, where m and n are respectively the degrees of polymerization of P1(PEG) and P2 (LC polymer).

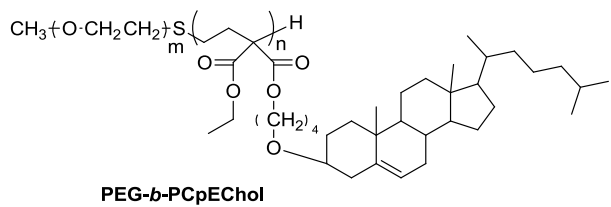

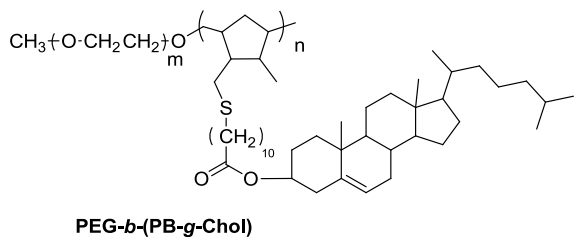

Scheme 2. Chemical structures of LC amphiphilic diblock copolymers synthesized by anionic ring-opening polymerization and thiol-ene polymer post-modification. The copolymers will also be noted as $P1_m$-b-$P2_n$, where m and n are respectively the degrees of polymerization of P1(PEG) and P2 (LC polymer).

The structural diversity of LC polymers is either introduced by different mesogens (Scheme 1), or by different polymer backbone. The mesogens can be simply thermotropic or photo-responsive; they can be side-on or end-on type; they can be nematic, smectic or cholesteric. Polyacrylate and polymethacrylate are the most common LC polymer backbone. Controlled radical polymerizations such as atom transfer radical polymerization (ATRP) were used to synthesize the LC copolymers from LC acrylate or methacrylate monomers. Yang et al., (9) (30) Xu et al. (32) and Jia et al. (10) (33) have prepared by ATRP LC amphiphilic block copolymers with poly(meth)acrylate as LC polymer backbone: PEG-b-PA444, PEG-b-PMAazoA444, PEG-b-PA6ester1 and PEG-b-PAChol (see Scheme 1). Copolymers with different hydrophilic/hydrophobic ratios (from 10/90 to 50/50) and different PEG and LC block lengths (e.g., molecular weight of 2000 Da and 5000 Da for PEG) were prepared. LC poly(meth)acrylates have a dense mesogen distribution along the backbone, *i.e.*, there is one mesogen every second consecutive carbons in the backbone. LC copolymers with lower mesogen densities in LC block were also been studied. LC copolymers with a polycyclopropane LC backbone whose repeating unit contains three consecutive carbons were synthesized by anionic ring-opening polymerization of cyclopropane-based monomer catalysed by phosphazene base (*t*-BuP$_4$) (see Scheme 2. PEG-b-PCpEChol). (34) These LC copolymers have one mesogen every third carbons in the LC block. LC copolymers with even less dense mesogens were further prepared, by grafting mesogens to PEG-*b*-PB (90% of 1,2-olefin) using thiol-ene reaction (see Scheme 2. PEG-*b*-(PB-*g*-Chol)).(39) In this case, 69% of 1,2-olefins in each chain were substituted by mesogenic side groups accompanied by backbone cyclization. PEG-*b*-(PB-*g*-Chol) polymers have approximately one mesogen every fourth carbons in the LC backbone.

The liquid crystalline properties of the LC homopolymers PA444, PMAazo444, PAChol, PA6ester1, PCpEChol and PB-*g*-Chol are listed in Table 1 together with one example of their amphiphilic block copolymers. PA444 and PMAazo444 are nematic polymers; (40), (41) PAChol, PCpEChol and PB-*g*-Chol are smectic polymers; (10), (34), (39) PA6ester1 presents both nematic and smectic phases. (32) In general, amphiphilic block copolymers preserve the mesophases of LC polymer block, but with lower transition temperatures in the case of similar molecular weight of LC polymer. As a matter of fact, the presence of PEG can result in microphase separation of block copolymers: e.g., PEG-*b*-PA6ester1 can self-assemble into hexagonal cylindrical organization where PEG cylinders are surrounded by the nematic or smectic matrix. (32) We will not discuss, in this paper, these self-assemblies of LC block copolymers in pure state. We just notice that the PEG and LC polymer block may have a limited miscibility. PEG can play the role of impurity in the absence of microphase separation, and even in the presence of microphase separation the intermaterial dividing surface may not be so clear-cut. Consequently, PEG-*b*-(LC polymer) has generally a lower transition temperature than the LC polymer alone.

The smectic structures of smectic polymers and copolymers are summarized in Table 2. The smectic A periods P were measured by small angle X-ray scattering (SAXS). In the homopolymer PA6ester1 and the copolymer PEG-*b*-PA6ester1, P = 2.52 nm and 2.37 nm, respectively. The extended molecular length of the side group is here estimated to be 2.35–2.45 nm by Dreiding models. A one-layer anti-parallel packing is therefore the most probable arrangement for the smectic phase (noted as $SmA_1$). (32) Similarly, PB-*g*-Chol and PEG-*b*-(PB-*g*-Chol) present also a $SmA_1$ with a one-layer anti-parallel packing, since the smectic period (3.58 nm and 3.75 nm) is close to the extended mesogen length $l$ = 4.60 nm. (39) The difference might come from the coiled conformation of the aliphatic spacer (longer than in PA6ester1). In contrast, PAChol and PEG-*b*-PAChol exhibit a smectic A phase with P = 4.30 nm and 4.25 nm, which correspond to a value between $l$ and $2l$, $l$ = 2.65 nm being the fully extended length of the cholesteryl mesogen, as estimated by Dreiding models. The smectic mesophase is then an interdigitated two-layer smectic A phase (SmA$_d$) with the side groups overlapping in the tails region. (10) (33) In a similar way, PCpEChol and PEG-*b*-PCpEChol present also an interdigitated two-layer smectic A phase (SmA$_d$). (34)

Table 1. Phase transition temperatures measured in LC homopolymer (P$_n$) and block copolymers (P1$_m$-*b*-P2$_n$) in pure state by DSC7 at 10°C.min$^{-1}$ and in polymersomes by N-DSC III at 1°C.min$^{-1}$ (taken as the peak temperatures in DSC thermograms on heating). m and n are respectively the degree of polymerization of P1 (PEG) and P2 (or P, LC polymer).

| Sample P$_n$ or P1$_m$-*b*-P2$_n$ (Hydrophilic/hydrophobic weight ratio) | Glass temperature T$_g$ (°C) | Smectic => Nematic transition T$_{SN}$ or Smectic => Isotropic transition T$_{SI}$ (°C) | Nematic => Isotropic transition T$_{NI}$ (°C) |
|---|---|---|---|
| PA444$_{24}$ | 55.6 | - | 115.1 |
| PEG$_{45}$-*b*-PA444$_7$ (30/70) | 30.3 | - | 59.3 |
| PEG$_{45}$-*b*-PA444$_7$ polymersomes | 27 | - | 84.8 |
| PMAazo444$_{167}$ | 57.4 [a,b] | - | 94.9 [b] |
| PEG$_{45}$-*b*-PMAazo444$_{12}$ (22/78) | 47.1 [a] | - | 68.4 |
| PEG$_{45}$-*b*-PMAazo444$_{12}$ polymersomes | 46 | - | 79.4 |
| PA6ester1$_8$ | 20 | 79.9 | 105.4 |
| PEG$_{45}$-*b*-PA6ester1$_{20}$ (20/80) | 15 | 60.8 | 73.9 |
| PEG$_{45}$-*b*-PA6ester1$_{20}$ polymersomes | 10 | 72.5 | 90.4 |
| PAChol$_{10}$ | 68 | 156.0 | - |
| PEG$_{45}$-*b*-PAChol$_{10}$ (28/72) | 45 | 115.9 | - |
| PCpEChol$_{13}$ | 33 [a,b] | 118.9 [a,b] | 122.0 [a,b,e] |
| PEG$_{45}$-*b*-PCpEChol$_{12}$ (23/77) | - | 160.0 [d] | - |
| PB$_{56}$-*g*-Chol | 35 [c] | 71.0 | - |
| PEG$_{91}$-*b*-(PB$_{33}$-*g*-Chol) (23/77) | 51 [c] | 106.0 | - |

[a] Determined from the cooling scan; [b] Measured at 5°C.min$^{-1}$; [c] Melting point (crystalline – smectic phase); [d] Measured by polarizing microscope upon heating at 1°C.min$^{-1}$; [e] A chiral nematic N* for PCpEChol$_{13}$.

Table 2. Smectic phases and smectic spacing P measured by SAXS for smectic homopolymers and copolymers

| Polymer[a] | Smectic phase | Smectic period P (nm) |
|---|---|---|
| PA6ester1$_8$ | SmA$_1$ | 2.52 |
| PEG$_{45}$-*b*-PA6ester1$_{20}$ (20/80) | SmA$_1$ | 2.37 |
| PAChol$_{10}$ | SmA$_d$ | 4.30 |
| PEG$_{45}$-*b*-PAChol$_{16}$ (20/80) | SmA$_d$ | 4.25 |
| PCpEChol$_{13}$ | SmA$_d$ | 4.45 |
| PEG$_{45}$-*b*-PCpEChol$_{12}$ (23/77) | SmA$_d$ | 4.58 |
| PB$_{56}$-*g*-Chol | SmA$_1$ | 3.58 |
| PEG$_{91}$-*b*-(PB$_{33}$-*g*-Chol) (23/77) | SmA$_1$ | 3.75 |

*2.2: Preparation of LC polymersomes*

Polymersome formation requires the mutual diffusion of water into the bulk block copolymer and vice versa.(42) In general, all reported methods for liposome preparation can also be used for polymersome formation. (43) (44) (45) In preparation protocols, the contact between water and polymer can be achieved directly or with the aid of organic solvent if the hydrophobic block is glassy at the preparation temperature. For block copolymers with hydrophobic blocks possessing a low $T_g$, such as PEG-*b*-PB ($T_{g,PB}$ ~ -90°C to -8°C according to their relative 1,4- and 1,2-olefin content), vesicles can be formed by direct hydration techniques assisted by sonication or electrical field (Dimova et al., 2002).(46) In contrast, block copolymers with a glassy hydrophobic block, such as PAA-*b*-PS ($T_{g,PS}$ ~ 100°C), often require an organic co-solvent to fluidize the polymer layers. Typically, a polymer solution is first prepared in an organic solvent and the solvent is then gradually exchanged with water.(47) This method belongs to a more general nanoprecipitation method based on the interfacial deposition due to the displacement of a solvent with the non-solvent. Recently, microfluidic and micro-patterning technology have opened some fascinating ways to prepare polymersomes with controlled size and efficient encapsulation.(48) (49) (50) (51)

Polymersome formation methods discussed above naturally lead to a symmetric copolymer distribution between both leaflets of bilayer membrane. One special method, called inverted emulsion, (45) permits to obtain asymmetric vesicles by independent assembly of the inner and outer leaflets of the vesicle.(52) Briefly, the inner monolayer is first formed via the emulsification of water droplets in oil containing the first amphiphile of interest. The outer monolayer is then formed by the centrifugation of the water droplets stabilized by the first amphiphile through the monolayer of the second amphiphile at the interface between a second oil solution (containing the second amphiphile of interest) and a water solution.

LC polymers present generally a $T_g$ higher than room temperature. Therefore, LC polymersomes were formed essentially by two methods: (1) nanoprecipitation for symmetrical nano-polymersomes; (2) inverted emulsion for asymmetrical or symmetrical giant polymersomes. In the method of nanoprecipitation, the choice of organic co-solvent depends on the structure of the LC polymer. The dioxane and the tetrahydrofuran (THF) that are miscible with water are the most commonly used co-solvent. Empirically, we found that most of LC copolymers studied with vesicle-forming hydrophilic/hydrophobic ratios (PEG-*b*-PA444, PEG-*b*-PMAazo444, PEG-*b*-PAChol, PEG-*b*-PA6ester1, PEG-*b*-PCpEChol) self-assemble into vesicles by using dioxane as co-solvent (but not with THF), (32) (33) (34) except the PEG-*b*-(PB-g-Chol) which can form vesicles only with THF as co-solvent (but not with dioxane). (39) In the inverted emulsion method, the organic solvent should be immiscible with and lighter than water. Vegetable oils are often used for liposomes. We have used toluene because of the good solubility of copolymer in it.(53)

## 2.3: LC polymer vesicles

### 2.3.1: Nematic polymer vesicles

A typical nanoprecipitation by adding water into the polymer solution in dioxane followed by a thorough dialysis again water was used for the polymersome preparation at room temperature. Well-structured unilamellar spherical polymersomes were obtained with nematic side-on LC copolymer PEG-*b*-PA444 and PEG-*b*-PMAazo444 in a rather large range of hydrophilic/hydrophobic weight ratios (e.g., from 40/60 to 19/81 in the case of PEG-*b*-PA444). (9), (30) Figure 3 shows cryo-electron micrographs of the polymersomes formed by $PEG_{45}$-*b*-$PA444_7$ (30/70) and $PEG_{45}$-*b*-$PMAazo444_{12}$ (22/78). The size of the observed vesicles is rather heterogeneous. Hydrodynamic diameters and distributions determined by dynamic light scattering (DLS) (see Table 3) confirm this size heterogeneity. Nevertheless, the membrane thickness is homogeneous: 10-11 nm for vesicles of $PEG_{45}$-*b*-$PA444_7$ and 14-15 nm for vesicles of $PEG_{45}$-*b*-$PMAazo444_{12}$ as measured by cryo-TEM for their hydrophobic part.

Table 3. Hydrodynamic diameters and distributions of LC polymersomes measured by DLS.

| Copolymers P1$_m$-b-P2$_n$ (hydrophilic/hydrophobic weight ratio) | Initial concentration in dioxane (wt%) | Hydrodynamic diameter (Z-average, nm) | PDI |
|---|---|---|---|
| PEG$_{45}$-b-PA444$_7$ (30/70) | 1% | 530 | 0.10 |
| PEG$_{45}$-b-PMAazo444$_{12}$ (22/78) | 1% | 820 | 0.19 |
| PEG$_{45}$-b-PA6ester1$_{20}$ (20/80) | 1% | 162 | 0.35 |

The thermotropic nematic nature of the LC membrane was studied by high sensitivity differential scanning calorimetry (N-DSC) on the polymersome dispersions in water.(54) Figure 4 shows the DSC thermograms of polymersomes of PEG$_{45}$-*b*-PA444$_7$ and PEG$_{45}$-*b*-PMAazo444$_{12}$ (first heating and cooling scans). Both copolymers in polymersomes exhibit a glass transition temperature, $T_g$, and a nematic–isotropic transition ($T_{NI}$) of the LC block. The transition temperatures are listed in Table 1, together with those of copolymers in the pure state (without water). For both copolymers, $T_{NI}$ is higher in vesicle than in pure state. This increase is expected, as the hydration of the PEG block increases the segregation between the two polymer blocks in the vesicle membrane, which should in turn enhance the nematic ordering of the mesogens. At room temperature where both polymersomes are formed, the hydrophobic blocks are in glassy nematic states because of their $T_g$ are higher than room temperature.

We discuss now the mesogen and chain organization of nematic polymer block in the polymersome membranes. (9) The PA444 block of PEG$_{45}$-*b*-PA444$_7$ has a degree of polymerization of 7 (monomer mass = 632 Dalton) and the PMAazo444 block of PEG$_{45}$-*b*-PMAazo444$_{12}$ a degree of polymerization of 12 (monomer mass = 602 Dalton). The mesogen length is 2.6 nm as measured by X-ray diffraction. (55) We consider that the backbone is extended and the mesogens are parallel to and distributed around the backbone, as revealed by small angle neutron scattering (SANS) studies.(56) (57) The LC hydrophobic block length is therefore estimated to be 4.4 nm for PA444 and 5.6 nm for PMAazo444. These values are close to half the values of the measured membrane thicknesses. In conclusion, membranes have a bilayer structure as shown in the schematic representation in Figure 3(b) and polymer vesicles are unilamellar. The mesogens have a radial organization in the spherical membrane. The LC backbones are rather stretched and present an elongated conformation similar to that of side-on nematic polymers in pure state.

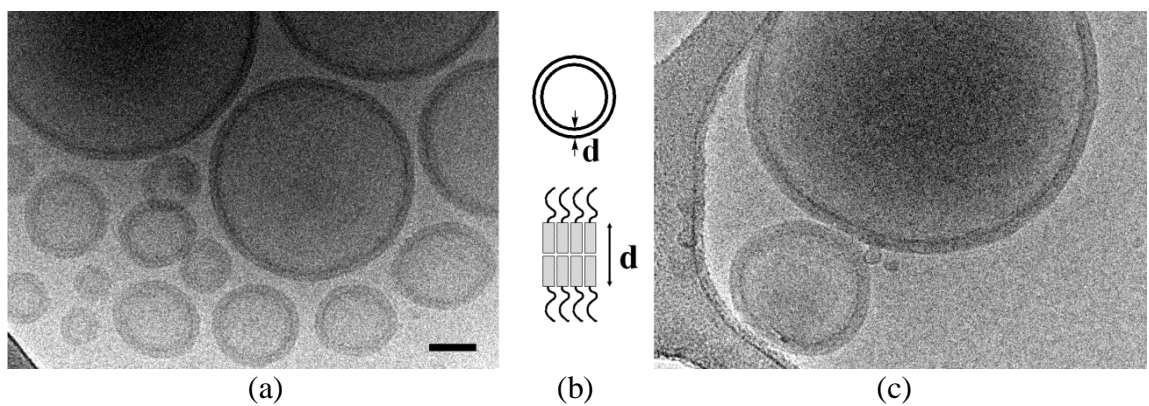

(a)          (b)          (c)

Figure 3. Cryo-electron micrographs of vesicles formed in water by PEG$_{45}$-$b$-PA444$_7$ (a) and by PEG$_{45}$-$b$-PMAazo444$_{12}$ (c). The scale bar at lower right of (a) is 50 nm and the scale is the same for (c). The mean lamellar thickness d is 10-11 nm for PEG$_{45}$-$b$-PA444$_7$ and 14-15 nm for PEG$_{45}$-$b$-PMAazo444$_{12}$. (b) is the schematic representation of diblock copolymers with bilayer structure in the membrane (the rectangles represent the LC blocks).

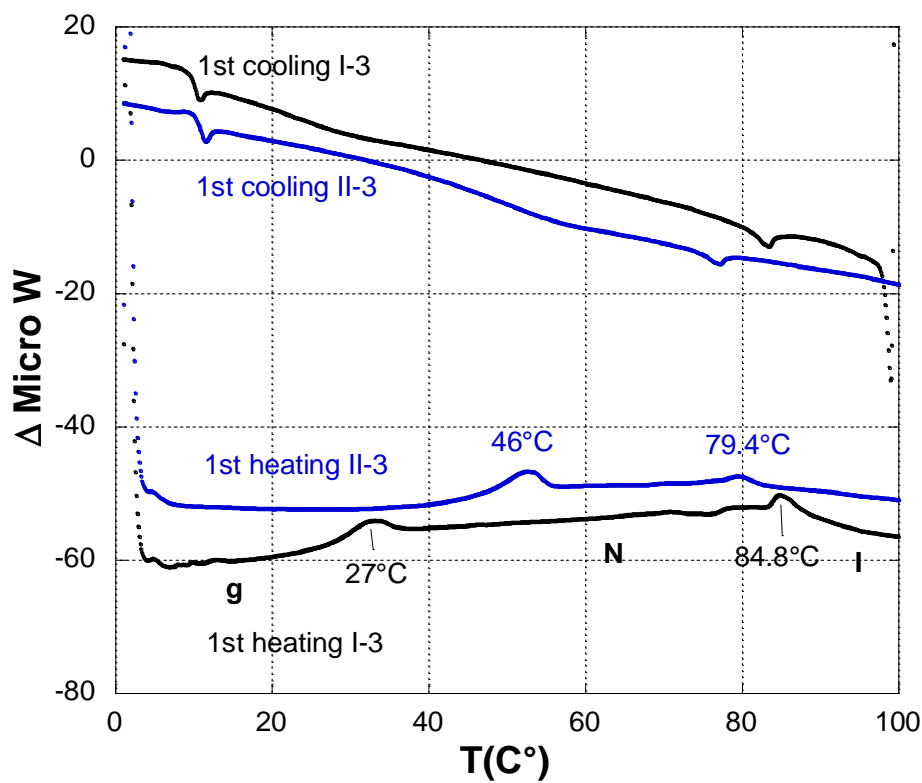

Figure 4. High sensitivity DSC thermograms of polymersome suspensions of PEG$_{45}$-$b$-PA444$_7$ (I-3) and of PEG$_{45}$-$b$-PMAazo444$_{12}$ (II-3).

*2.3.2: Smectic polymer vesicles*

Four families of amphiphilic block copolymers with smectic LC blocks, PEG-*b*-PA6ester1, PEG-*b*-PAChol, PEG-*b*-PCpEChol and PEG-*b*-(PB-g-Chol), have been used for the polymersome formation using typical nanoprecipitation method. What is striking in these polymersomes is that they are not spherical as classical polymersomes or nematic polymersomes discussed above.

*2.3.2.1: Faceted polymersomes.* $PEG_{45}$-b-$PA6ester1_{20}$ polymersomes are faceted. Figure 5a shows one of the numerous faceted vesicles observed in $PEG_{45}$-b-$PA6ester1_{20}$ polymersomes.(32) Periodic stripes are clearly visible perpendicular to the membrane surface. Their period is measured as 2.5 nm in agreement with the period of $SmA_1$ in the homopolymer PA6ester1 and in the copolymer PEG-b-PA6ester1. The membrane thickness is here again rather homogeneous (e = 10 nm) proving a rather stretched backbone in the bilayer membrane. Figure 5b is the schematic representation of the molecular organization of mesogens ($SmA_1$) and polymers in the membrane. N-DSC measurement performed on PEG-b-PA6ester1 polymersome dispersion (see Figure 6) do confirm the smectic nature of the vesicle membrane, the phase sequence being g-15°C-SmA-60.8°C-N-73.9°C-I (see also Table 1 for the comparison with pure copolymer).(58) The size of polymersomes is heterogeneous as shown by DLS measurement (see Table 3).

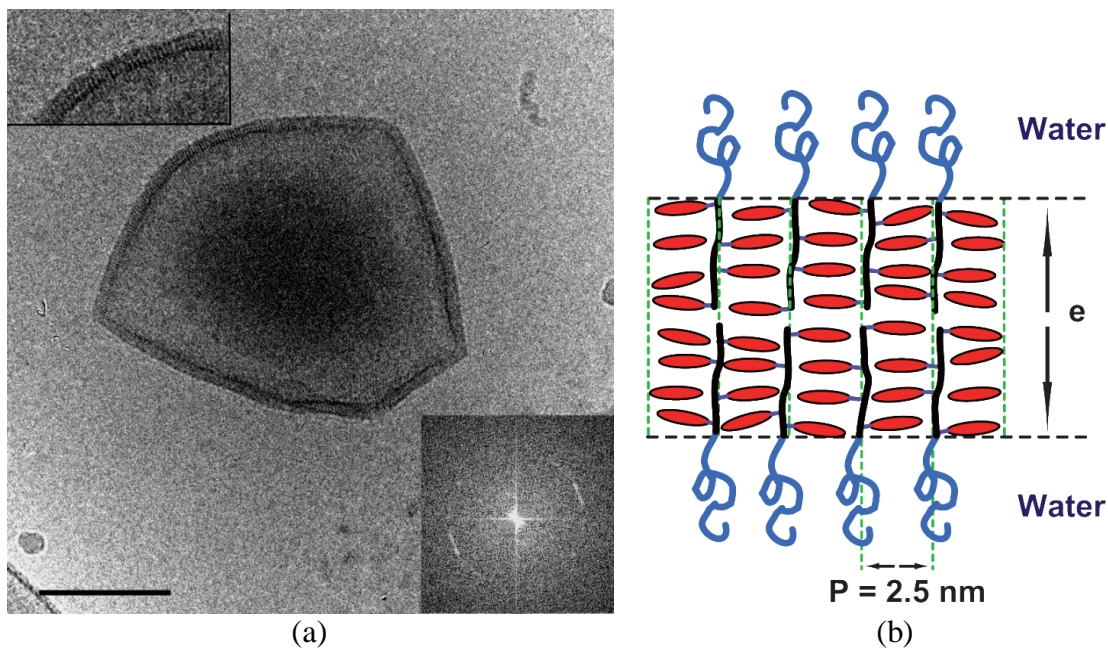

(a)                      (b)

Figure 5. (a) Cryo-transmission electron micrographs of polymer vesicles of $PEG_{45}$-$b$-$PA6ester1_{20}$. The inset at higher left is an enlargement of the upper left area of the vesicle in order to highlight the smectic stripes. The inset at lower right is the Fourier transform of the image, diffraction spots corresponding to a period of $P = 2.5 \pm 0.1$ nm. Scale bar =100 nm. (b) Schematic representation of the smectic molecular organisation within a cross section of the membrane of $PEG_{45}$-$b$-$PA6ester1_{20}$ polymersome. The hydrophilic PEG in blue connected to the hydrophobic side-chain LC polymer which itself consists of a black backbone and red LC mesogens represented by small elongated ellipsoids. The smectic A structure is a one-layer antiparallel packing ($SmA_1$). The membrane thickness e is about 10 nm (e is used to represent the thickness of smectic membrane, while d is used for nematic membrane as shown in Fig. 3).

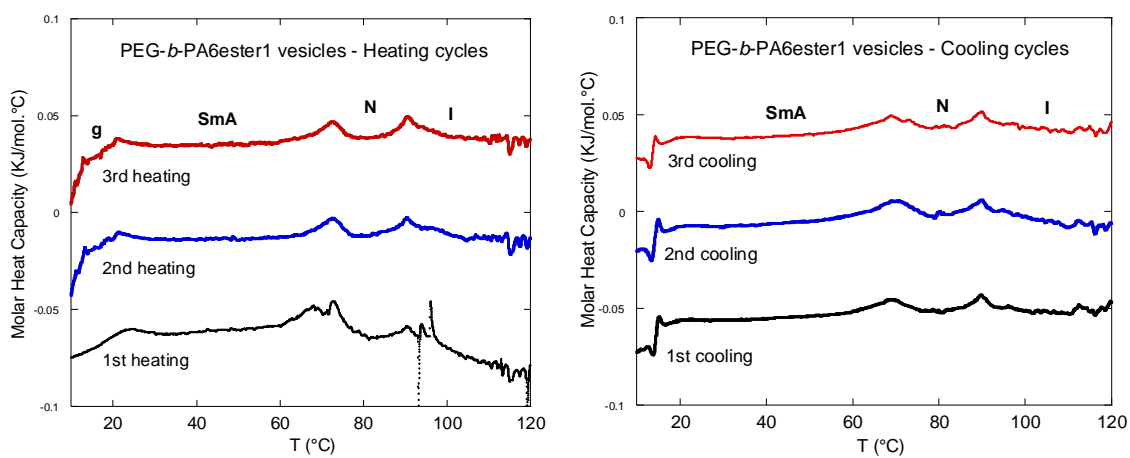

Figure 6. High resolution DSC curves of $PEG_{45}$-$b$-$PA6ester1_{20}$ polymersomes at 1°C min$^{-1}$. Molar heat capacity (KJ · mol$^{-1}$ · °C$^{-1}$) as a function of temperature. The parasite signals around 93°C at the first heating scan are electronic signals of apparatus. g represents the glassy state, SmA the smectic phase, N the nematic phase and I the isotropic phase. Curves have been shifted artificially along the ordinate for the sake of clarity.

*2.3.2.2: Ellipsoidal polymersomes.* The polymersomes of PEG$_{45}$-*b*-PAChol$_{10}$ and PEG$_{45}$-*b*-PAChol$_{16}$ are ellipsoidal.(10) Figure 7 shows their images observed by cryo-TEM. Some of them have spherical buds emanating from the poles (Figure 7b), and others have smooth tips (Figure 7a and 7b).

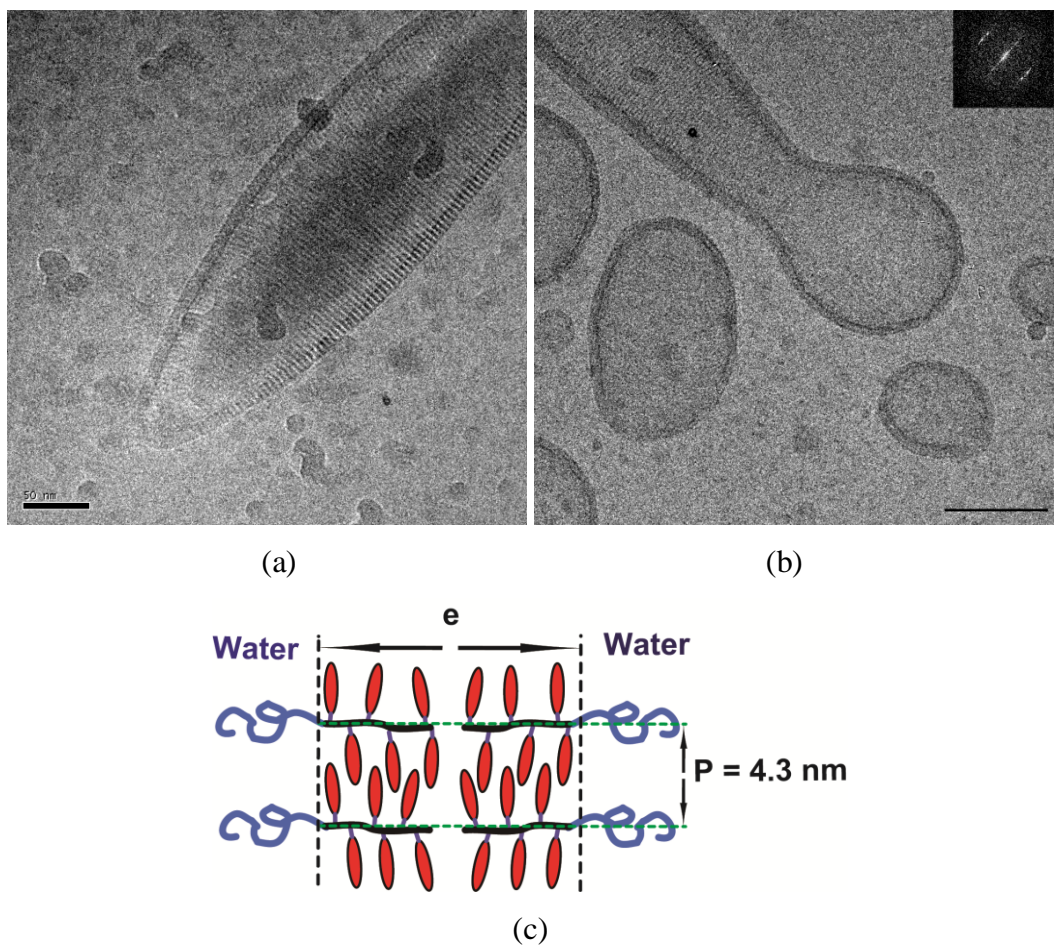

(a)          (b)

(c)

Figure 7. Cryo-transmission electron micrographs of smectic polymer vesicles of PEG-*b*-PAChol. (a) PEG$_{45}$-b-PAChol$_{16}$, scale bar = 50 nm. (b) PEG$_{45}$-b-PAChol$_{10}$, scale bar = 100 nm. Inset in (b) is Fourier transform of representative areas of the vesicles. The periodicity of all smectic areas is identical and corresponds to P = 4.3 ± 0.1 nm. The buds of the vesicles in (b) is not liquid crystalline as seen on the Fourier transform (not shown). (c) Schematic representation of the smectic molecular organisation within a cross section of the membrane of PEG-*b*-PAChol polymersome. See Figure 6 for the symbols of mesogens and polymer chains. The smectic A structure is a interdigitated two-layer packing (SmA$_d$). e is the membrane thickness.

PAChol exhibits only a SmA$_d$ phase with SmA$_d$-I transition higher than 100°C. Therefore N-DSC experiment on polymersome dispersion in water is not applicable to phase transition measurement. The smectic nature of polymersomes was then revealed directly by small angle X-ray scattering on polymersome dispersion sample. Figure 8a shows SAXS intensity profile. The diffraction peaks at 1.458 nm$^{-1}$ for both polymersomes give a smectic period of P = $2\pi/q_{max}$ = 4.30 nm, in agreement with that of PAChol in pure state. What is more interesting in these SAXS intensity profile is we can observe an oscillation at q ~ 0.5 nm$^{-1}$, that is the signature of membrane structure. In a plot of $q^2I(q)$ versus q, the first minimum corresponds then to the first zero of the membrane form factor. This gives the value of membrane thickness $e_{total}$ = $2\pi/q_{min}$, which is the total thickness including PEG part and PAChol part. We found $e_{total}$ = 22.7 nm ($q_{min}$ = 0.2772 nm$^{-1}$) for PEG$_{45}$-b-PAChol$_{10}$ and $e_{total}$ = 23.2 nm ($q_{min}$ = 0.2706 nm$^{-1}$).

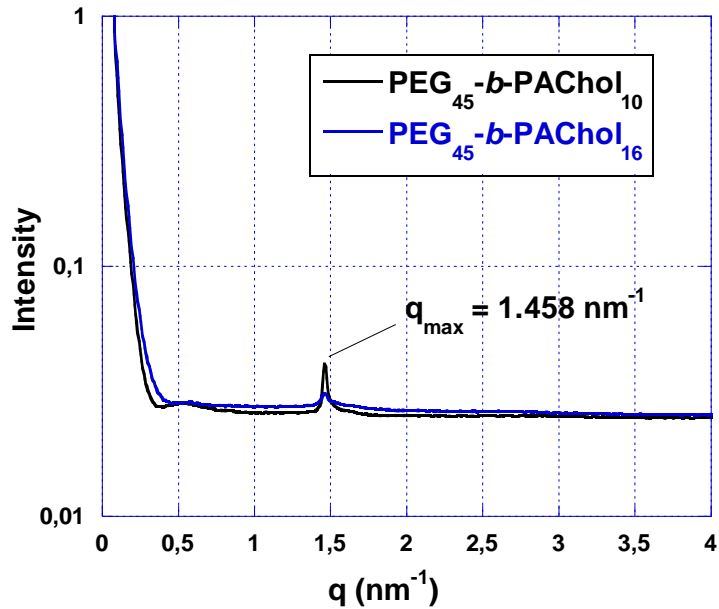

Figure 8. SAXS intensity profile of PEG-*b*-PAChol polymersome dispersions in water. Measurements were made at Swing line in SOLEIL synchrotron center. Narrow peaks ($q_{max}$ = 1.458 nm$^{-1}$) correspond to the smectic order of 4.30 nm. An oscillation at q ~ 0.5 nm$^{-1}$ is also observed for each curve that is the signature of membrane structure (see main text).

We return now to the cryo-TEM images for membrane structure details. Periodic stripes are clearly present in the most part of the membrane except in the two extremities of the polymersomes (spherical buds or smooth tips are isotropic). The period measured is nothing else than P = 4.30 nm, the smectic layer spacing of PAChol in pure state and in polymersomes. Figure 7c shows the schematic representation of the smectic molecular organisation (SmA$_d$) within a cross section of the membrane. The membrane thickness is of considerable interest and turns out to be substantially different in smectic regions and isotropic regions. For $PEG_{45}$-$b$-$PAChol_{10}$, the thickness falls in the range e = 8 – 13 nm in smectic regions but is thinner in the buds where no stripes are visible (in the range $e_I$ = 4 – 7 nm). For $PEG_{45}$-$b$-$PAChol_{16}$, the corresponding thicknesses are e = 10 – 13 nm and $e_I$ = 5 – 7 nm respectively. Note that cryo-TEM is only sensitive to the hydrophobic part of the membrane. This is why these thicknesses are smaller than those measured by SAXS which is sensitive to both PEG and LC blocks. The hydrophobic thickness of the membrane here is very similar for both copolymers in spite of the fact that the LC block chain length of $PEG_{45}$-$b$-$PAChol_{16}$ (long LC block with n = 16) is on average 60% longer than that of $PEG_{45}$-$b$-$PAChol_{10}$ (short LC block with n = 10). The conformation of hydrophobic chains must therefore be very different in each case. Also note that bending of the membrane parallel to the smectic layer doesn't change the layer spacing, while bending perpendicular to the layers makes the layer spacing on both sides of the membrane unequal, and therefore costs extra elastic free energy. This explains why the stripes are always perpendicular to the major axis of vesicles (see Figure 9 for the illustration of smectic ellipsoidal polymersomes). Because the LC polymer membranes possess polymer and liquid crystal characteristics in the same time, two elastic energies should be considered: one is the Frank elastic energy of liquid crystals and another is the bending energy of the polymer membrane. Bending a membrane with longer polymer chains and with more mesogens should be more energy-cost than for a membrane with shorter polymer chains and less mesogens. This is why $PEG_{45}$-$b$-$PAChol_{16}$ vesicles are on average larger (smaller membrane curvature) than $PEG_{45}$-$b$-$PAChol_{10}$ vesicles. For an even longer copolymer, $PEG_{114}$-$b$-$PAChol_{60}$, the bending perpendicular to the layers becomes impossible in some conditions. Consequently longer nanotubes instead of ellipsoidal vesicles were obtained (detail see ref.(11)).

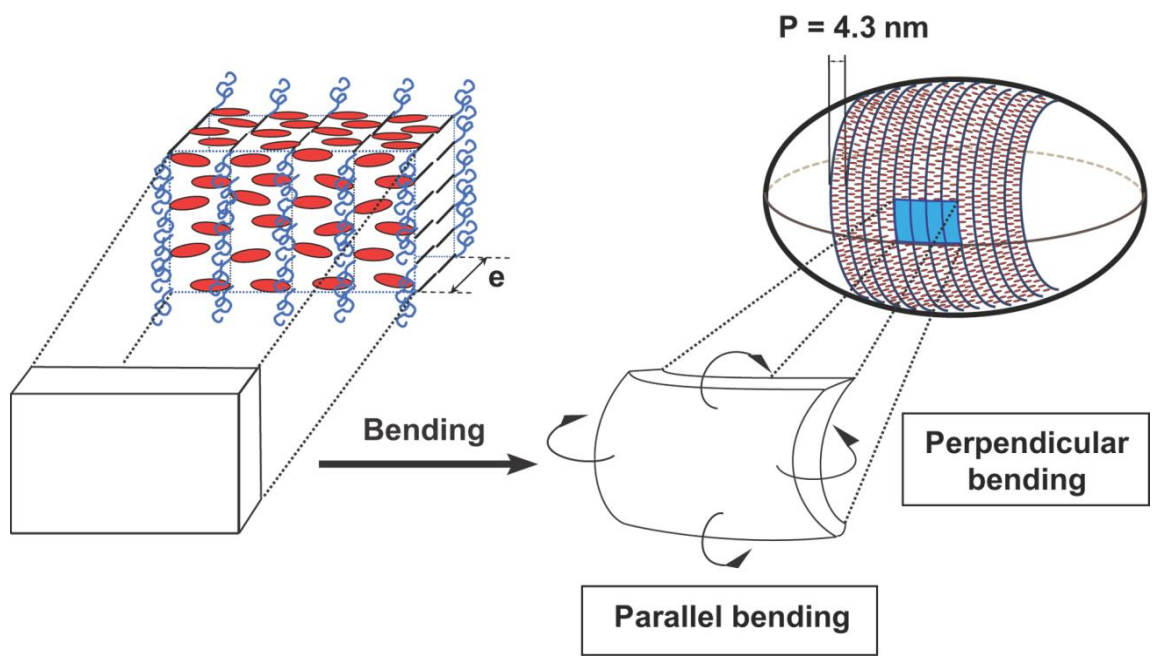

Figure 9. Schematic representation of a smectic ellipsoidal polymer vesicle. See Figure 6 for the symbols of mesogens and polymer chains. e is the membrane thickness and P is the smectic period.

*2.3.2.3: Theoretical consideration.* The smectic polymersomes exhibit two-dimensional smectic order. (10) If only this LC property is considered, striped, or smectic, order on a surface with spherical topology must exhibit orientational defects of total charge +2,(59) (60) (61) as required by the Gauss-Bonnet-Poincaré theorem. For the polymer vesicles discussed here, since all the smectic layers are roughly perpendicular to the major axis and each polar region should carry disclination charge +1. Moreover it is proposed that all stripes should generically form helices around the major axis, because the special case where all smectic layers form closed circles is possible but unlikely. Consequently, the +1 disclinations around each pole are more appropriately characterized as tightly bound pairs of +1/2 disclinations, as illustrated in Figure 10. These topological defects could also be the nucleator of vesicle budding and result in spherical isotropic buds that relieve also topological constrain, as shown in Figure 7b. The budding phenomenon is, therefore, a result of smectic-isotropic phase separation (See ref.(10) for the theoretical analysis of the necks of buds). This phase separation must happen during the removal of dioxane, since in the absence of the fluidizer at room temperature the system is in a glassy state where phase separation is not possible.

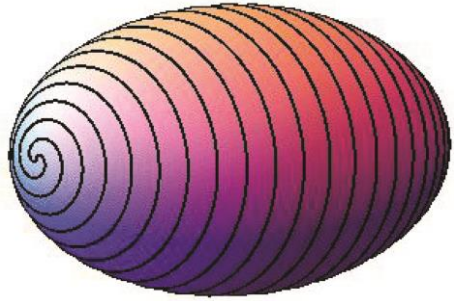

Figure 10. Expected defects structure at a pole of ellipsoid of a smectic ellipsoidal polymer vesicle.

Xing et al. have developed further a more complete theoretical model for the morphology of the membrane structure with internal nematic/smectic order.(62) The morphology of a bilayer is controlled by the competition between the bending energy of polymer bilayer and the Frank free energy of LC structure. Using both analytic and numerical approaches, it was shown that the possible low free energy morphologies include nano-size cylindrical micelles (nano-fibers), faceted tetrahedral vesicles, and ellipsoidal vesicles, as well as cylindrical vesicles (nano-tubes). Two limiting cases were considered: $K \ll \kappa$ and $K \gg \kappa$ ($K$ is the Frank constant and $\kappa$ the bending rigidity). In the case of $K \ll \kappa$, the dominant contribution to the total energy is then the bending energy: minimizing this leads to a round spherical shape. However, as the bending energy is not isotropic, the shape will reflect the anisotropy of the bending moduli, leading to ellipsoidal shapes, as observed experimentally and above-discussed. In the case of $K \gg \kappa$, the system should first minimize the Frank free energy: the ground state morphology of a vesicle with spherical topology is a faceted tetrahedron, with a strength 1⁄2 disclination located at each of the four corners. This structure is indeed observed in our experiments, (11) (32) as well as in the simulation (Figure 11).

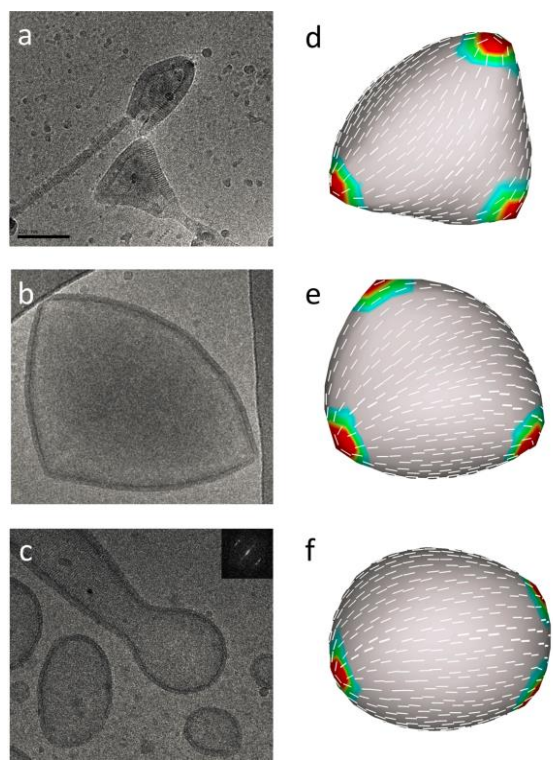

Figure 11. Comparison between experimental observations (a)–(c) and computer simulations (d)–(f). Left: Experimental results, (a) a tetrahedral smectic vesicle (20); (b) a fat tetrahedral smectic vesicle (15); (c) an ellipsoidal smectic vesicle (14). Right: Simulation results for a case of $K_3/K_1 \approx 2.0$, (d) $\kappa = 0.04$; (e) $\kappa = 0.1$; (f) $\kappa = 0.5$. The contour plots show the distribution of the local Frank free energy. $K_3$ and $K_1$ are splay and bend Frank constant, respectively.

*2.3.2.4: Other shapes of smectic polymersomes.*

In the LC block copolymer, the structural diversity can be introduced not only by the different LC orders, but also by the different backbones. PEG-*b*-PCpEChol and PEG-*b*-(PB-g-Chol) with polycyclopropane and polybutadiene-based LC blocks have also been studied (see *2.1: LC amphiphilic block copolymer*s). The flexibility is polyacrylate < polycyclopropane < polybutadiene-based chain and the mesogen density along the backbone is polyacrylate > polycyclopropane > polybutadiene-based chain. Consequently, PEG-*b*-PCpEChol and PEG-*b*-(PB-g-Chol) form smectic polymersomes with more flexible membranes.

Figure 12a shows a polymersome with wavy membrane for PEG-*b*-PCpEChol, (34) and Figure 12b shows spherical and multilayer polymersomes for PEG-*b*-(PB-g-Chol).(34) In these membranes, cholesteryl mesogens are organised again in a smectic structure with layer normal (and mesogen director) parallel to the polymer bilayer: a SmA$_d$ phase with P = 4.7 nm for PEG-*b*-PCpEChol and a SmA$_1$ phase with P = 3.9 nm for PEG-*b*-(PB-g-Chol). The fact that the bilayer membrane of the PEG$_{45}$-*b*-PCpEChol polymersome twists or waves (without apparent periodicity) is basically because of the low elastic energy of membrane. As for PEG-*b*-(PB-g-Chol), one of the remarkable features of the polymersomes is they are often not unilamellar. Most of them are striped hollow concentric spherical vesicles or eccentric complex vesicles (multi-lamellar). On the one hand, self-assembly in water of other PEG-*b*-smectic polymers had produced, among other aggregates, faceted or ellipsoidal or wavy unilamellar striped vesicles. On the other hand, hollow concentric vesicles have been found for classical amphiphilic block copolymers such as PS-*b*-PAA, PEO-*b*-PBA and PEO-*b*-PBO.(63) (64) (65) (66) The pronounced flexibility of PB backbone as compared to polyacrylate backbone, and its incomplete functionalization by the mesogenic side groups (only 69% substitution), might confer to the membrane the flexibility needed to give the vesicle the possibility to adopt a spherical shape and to form multilamellar organization. In order to better understand what parameters govern these morphologies, further experiments are needed, including the synthesis of new diblock copolymers with diversified macromolecular backbones and/or mesogenic groups.

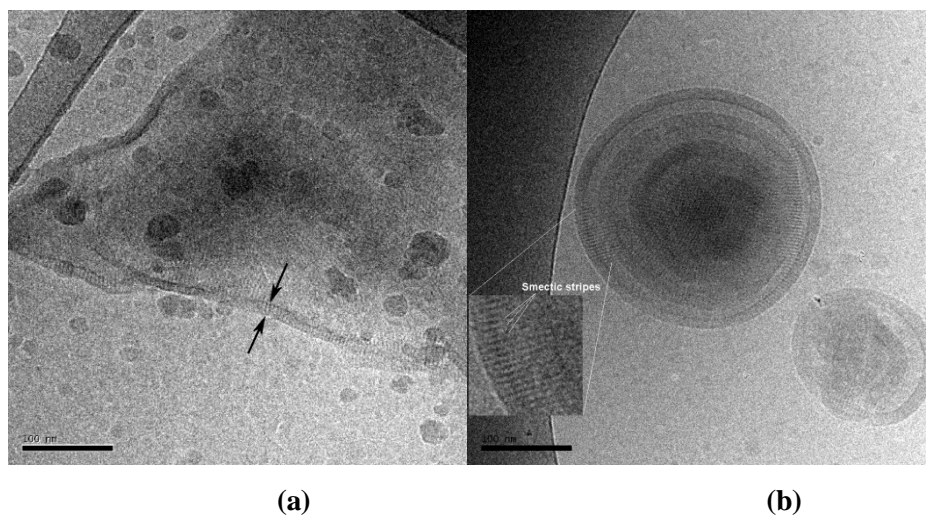

(a) (b)

Figure 12. (a) Vesicular structure with wavy membrane formed by PEG$_{45}$-*b*-PCpEChol$_{12}$. The stripes in cryo-electron micrographs have a periodic spacing P = 4.7 nm. The membrane thickness e = 11-14 nm. Scale bar = 100 nm. (b) Complex multilayer vesicles formed by PEG$_{91}$-*b*-(PB$_{33}$-*g*-Chol). Cryo-TEM images show also clearly the smectic stripes of P = 3.9 nm. The membrane thickness is e =13 - 14 nm.

## 3. Photo- and thermo-responsive liquid crystalline polymersomes

We discuss now the use of the LC copolymers with photo-induced phase transitions and of the LC copolymers with transition temperatures lower than 100°C (boiling point of water) to design photo- and thermo-responsive LC polymersomes. We have concluded from the structural studies of nematic polymersomes that the side-on nematic polymer chains have an elongated conformation perpendicular to the membrane and the mesogens are organized in a radial manner in the spherical bilayer (see Figure 13a). (9) On the other hand, from the previous studies of LC polymer conformation by SANS (56) (57) and the studies of thermo- and photo-responsive LC polymer actuators,(55) (67) we got to know that the side-on nematic polymers undergo conformational changes from elongated to spherical shape at $T_{NI}$ transition either trigger by temperature change or by *trans-cis* photo-isomerization of azo-mesogen under UV illumination (see Figure 13b). The obvious consequence of this conformational change is the size contraction along the mesogen orientation. We describe in the following how to take advantage of these conformational changes to trigger the vesicle opening.

*3.1: Photo-responsive LC polymersomes*

Giant polymersomes (> few microns in diameter) were prepared by the method of inverted emulsion from azobenzene-containing block copolymer PEG-*b*-PMAazoA444, in order to follow the photo-responsive morphological changes of polymersomes under optical microscope.(12) Figure 14 shows the morphological evolution of PEG-*b*-PMAazoA444 polymersome under UV illumination. After 300 s of illumination, the polymersome doesn't seem to open, but the membrane wrinkles. The membrane wrinkling can be explained by the increase of its surface area. Starting from a thin, cigar-like shape corresponding to N state, UV irradiation transforms the LC hydrophobic block to a coil characterized by a contraction of LC blocks perpendicular to the bilayer and an increased molecular area along the bilayer (see Figure 13, it is reasonable to suppose the polymer volume doesn't change).

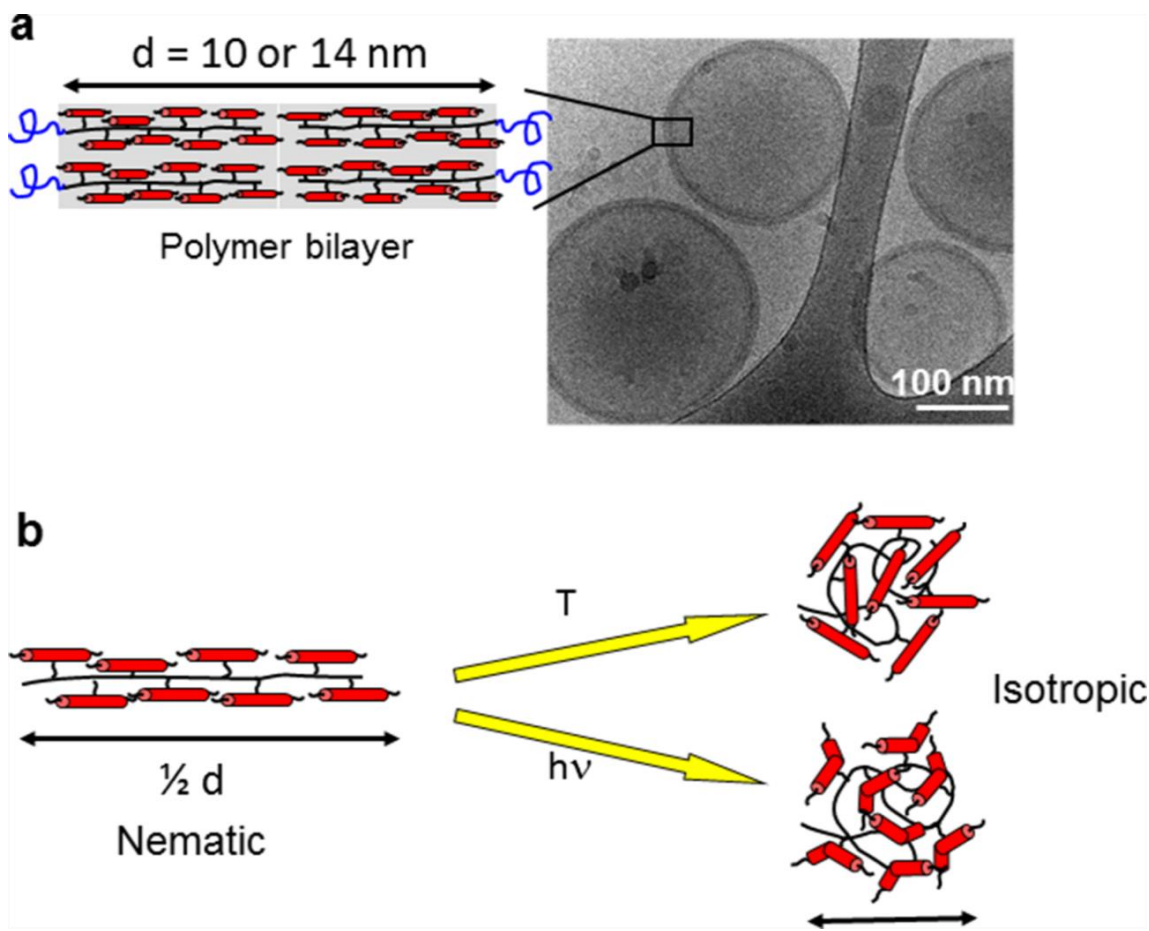

Figure 13. (a) Schematic representation of the molecular organization of side-on nematic polymers in the bilayer membrane of polymersomes of PEG-*b*-PA444 and PEG-*b*-PMAazo444 (cryo-TEM of polymersomes in the right). (b) The conformational change from a cigar-like shape to a coil of the side-on nematic polymer induced by a nematic-isotropic (N-I) phase transition. This N-I transition can be triggered either by a temperature variation (T) or by a photo-chemical isomerization of azobenzene under UV illumination (hν).

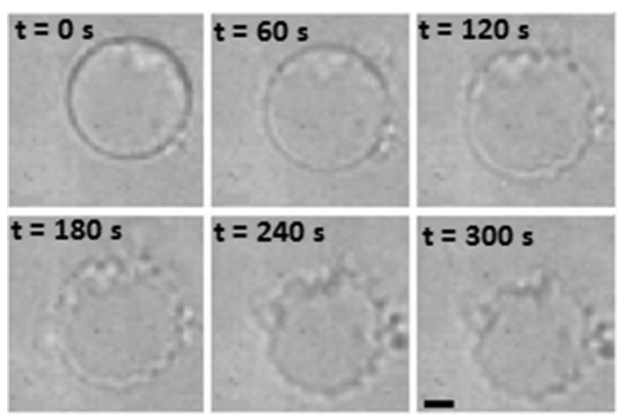

Figure 14. Membrane wrinkling of a PEG-*b*-PMAazo444 giant polymersome under UV illumination. Scale bar = 10 μm.

The basic idea is then to introduce frustration in the polymersome membrane by breaking up the bilayer symmetry. To implement this approach, we prepared asymmetric polymersomes in which each leaflet consisted of a different type of diblock copolymer: one copolymer was insensitive to any remote stimulus (PEG-*b*-PB, or PBD for simplicity), while the hydrophobic moiety of the second copolymer was the light sensitive LC polymer PEG-*b*-PMAazo444 (PAzo).(12) Figure 15a & b shows the chemical structures of the two selected copolymers and a cartoon of the LC copolymer conformation in the membrane both in the absence and in the presence of UV light for polymersomes ePBD-iPAzo (external leaflet = PBD, inner leaflet = PAzo). UV illumination leads to an increased molecular area of inner PAzo leaflet, while the area of external PB leaflet doesn't change. Consequently, the net effect in the mesoscopic scale is the creation of spontaneous curvature of the membrane, which triggers membrane rupture and polymersomes bursting. Indeed, exposure to UV illumination around 360 nm caused vesicle rupture which was completed in less than a few hundreds of milliseconds and the release of the substance from the interior compartment (Figure 16). Rapid pore opening (probably via heterogeneous pore nucleation, *e.g.*, on a defect) was accompanied by the formation of an outward curling rim, as expected to be generated by the change of spontaneous curvature in the membrane where the inner leaflet is light-responsive (Figure 15c). The polymersome bursting takes place also if the inner leaflet is inert and the external leaflet is light-responsive, but with inward curling rim during the vesicle opening. These results highlight a new general strategy to create stimuli-responsive polymersomes based on the fabrication of asymmetric membranes, and driven by a change in membrane spontaneous curvature.

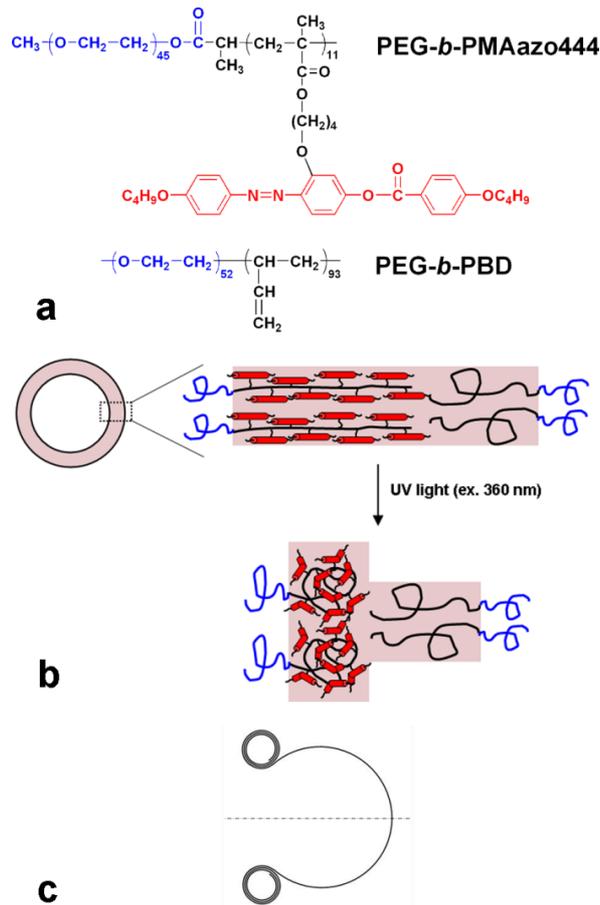

Figure 15. Copolymers and bilayer conformation. (a) Chemical structures of copolymers, PEG-*b*-PBD and PEG-*b*-PMAzo444. (b) Cartoon of a polymersome and cartoon depicting the conformation of both copolymers within the bilayer for an ePBD-iPAzo vesicle. The PEG-*b*-PBD is always in a coil-coil state. In the absence of UV light, the hydrophobic LC block of the PEG-*b*-PMAzo444 has a rod-like conformation (corresponding to a nematic state). Under UV illumination, isomerization of the mesogenic groups induces a conformational change of the polymer backbone to a disordered isotropic state. The net effect of UV exposure is two-fold: at the molecular scale, the projected area of the LC block is increased; at the mesoscopic scale, the spontaneous curvature of the bilayer is increased. (c) Schematic representation of pore opening driven by outward curling (for ePBD-iPAzo).

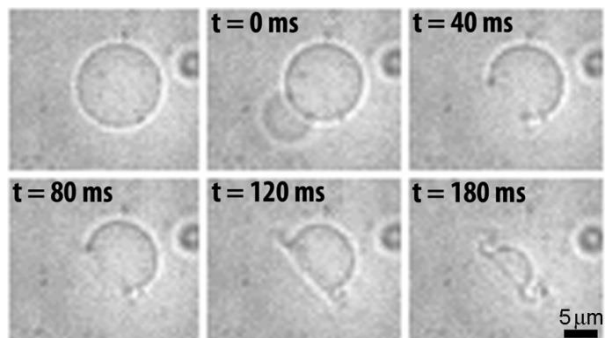

Figure 16. Snapshots of an asymmetrical ePBD-iPAzo polymersome bursting under UV illumination. Bright-field images were taken using a high-speed digital camera. The first image shows the vesicle prior to illumination. Time t=0 corresponds to pore nucleation. The expulsion of sucrose solution is visible as the pore nucleated. The other images correspond to pore growth and clearly show outward spirals (scale bar = 5 μm).

## 3.2: Thermo-responsive LC polymersomes

The hydrophobic blocks in the polymersomes of PEG-*b*-PA444, PEG-*b*-PMAazo444 and PEG-*b*-PA6ester1 are nematic or smectic polymers displaying LC phase transition temperatures lower than 100°C, (9,68) (32) and the hydrophilic block is a PEG of 2000 Da. Of course, we wanted to study the effect of the thermotropic LC phase transitions on the structure of polymersomes. However, the surprising finding is that the structural changes of polymersomes are driven in all cases by the critical dehydration of PEG corona (at around 55°C), but not by the LC transition temperature. PEG is quite a peculiar polymer. It is water soluble at room temperature, but displays a LCST (low critical solubilization transition) behavior with transitions higher than the boiling point of water (100°C). However, a lot of studies showed that the solubility (or degree of hydration) of PEG decreases already with increasing temperature in the range below this LCST,(69) especially when PEG chains form the hydrophilic corona of colloids.(70)

LC polymersomes display drastic and irreversible structural changes when heated above ~55°C, as revealed by the studies cryo-TEM and SANS. These changes are not influenced by the LC transitions (nematic-isotropic and smectic-nematic), but are dependent on the LC structures and membrane mechanical properties. Nematic polymersomes turn into capsules with a thick wall containing dehydrated PEG chains, whereas smectic polymersomes collapse into dense aggregates with microphase separation (see Figure 17). A decrease or disappearance of the volume of inner aqueous compartment occurs in these drastic and irreversible structural changes, which can be used for thermally controlled release, as shown in our studies of calcein release.(58) Even though the critical temperatures of around 55°C, higher than *in vivo* physiological temperature (~37°C), don't allow the *in vivo* use of the systems, we could envision possible *in vitro* biotechnological applications, for example, as nanoreactors, which release products after reaction at a still mild temperature (≤ 75°C).

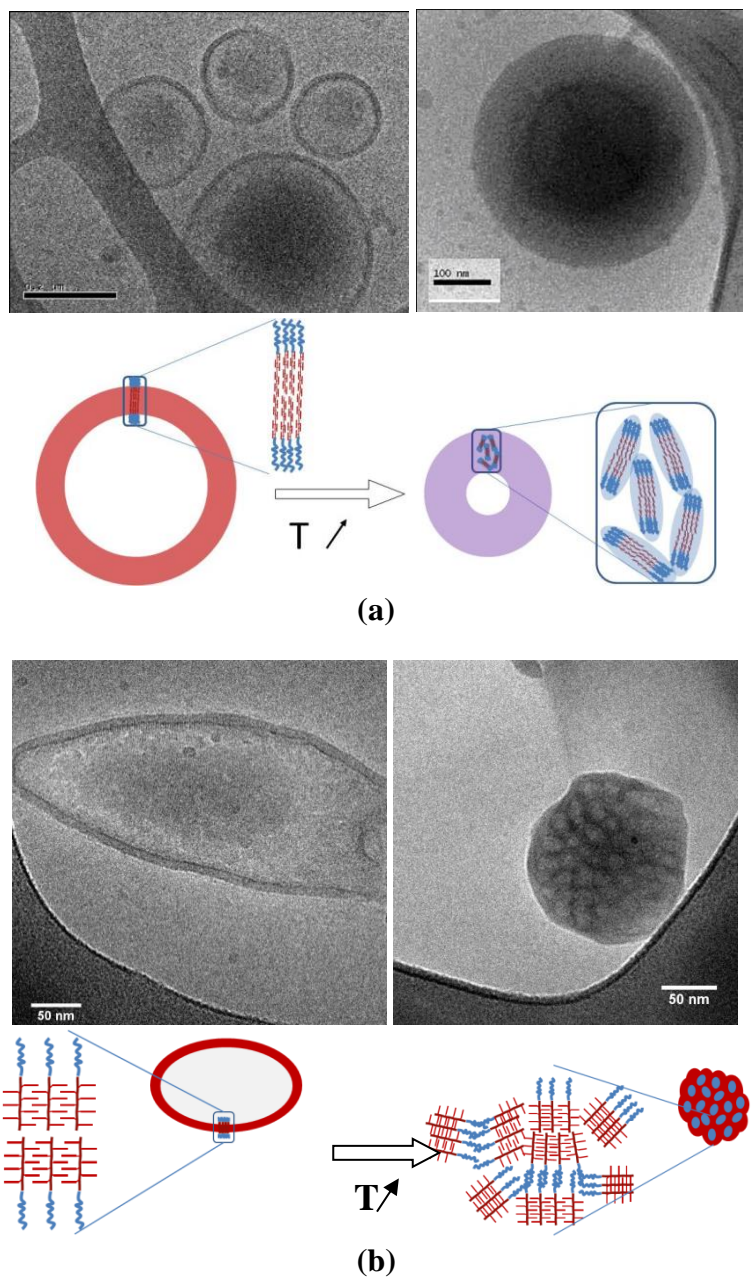

Figure 17: Thermally induced morphological evolutions of LC polymersomes revealed by Cryo-TEM images. (a) Nematic polymersomes made from copolymer $PEG_{45}$-*b*-$PMAazo444_{12}$ (upper left, scale bar = 200 nm) with their structure cartoon (lower left), and thick-walled capsules obtained after heating to 90°C (upper right, scale bar = 100 nm) with their structure cartoon (lower right). (b) Smectic polymersomes made from copolymer $PEG_{45}$-*b*-$PA6ester1_{20}$ (upper left, scale bar = 50 nm) with their structure cartoon (lower left), and microphase-separated dense nano-objects obtained after heating to 75°C (upper right, scale bar = 50 nm) with their structure cartoon (lower right). In the cartoons, blue parts represent PEG chains and red parts LC polymers.

## 4. Concluding remarks

We have described in this review the lyotropic polymer vesicles exhibiting thermotropic nematic and smectic LC structures in their hydrophobic membranes. We have shown how the morphology of polymersomes can vary with the structure of LC hydrophobic polymer block. The remarkable result is the observation of ellipsoidal and tetrahedral polymer vesicles. We discussed also the design of stimuli-responsive polymersomes after getting the precise knowledge of the molecular organisation of mesogens and of chain conformation in the membrane.

The biological world exhibits membranes with a wide variety of shapes and orders (tubules, buds and faceted structures). Tubules and buds are two important examples of shapes found in cellular and subcellular processes and their formation has been investigated theoretically(71)(59)(72)(73) and experimentally.(74) It has been shown, for example, that the coupling of the molecular tilt order of the lipids (relative to the normal of the membrane) to the Gaussian curvature favors cylindrical shapes for vesicles. (71) (73) (74) Another theory proposed, however, that tubules could be formed from spherical liposomes without this chiral interaction. The transformation requires only the development of orientational order within the bilayer.(59) In this paper the smectic order in the polymer membrane of ellipsoidal towards tubular vesicles was directly imaged, and isotropic buds were observed at the extremities of ellipsoidal polymer vesicles. Faceted surface structures were studied previously in large viral capsids,(75)(76) which are formed by crystalline packing of proteins. There the faceting is energetically favorable because it reduces the in-plane strain energy of the crystalline order formed by the constituent proteins. What we have shown in this paper is that a similar faceting can also be driven by the Frank free energy of LC order, despite their "liquid" crystalline nature. The experimental and theoretical studies of these fascinating morphologies of nematic/smectic polymer vesicles could pave the way for formulating guiding principles in designing nano-carriers of active substances with specific shapes. As the ellipsoidal and tetrahedral smectic vesicles possess topological defects (bi-poles and four corners), we speculate that these vesicles could also allow creation of novel divalent and tetravalent colloids with ligands or other functional groups anchored at the defect cores.(61)(77)

The combination of two soft matters (liquid crystals and polymers) resulted in a special molecular organization of side-on nematic polymer in the vesicle membrane.

Using an asymmetrical design of the polymersome bilayer, the side-on nematic polymer forming only one leaflet can act as a nano-actuator in the opening of asymmetrical polymersomes, the motor of the actuation is the conformational change of LC polymers in the N-I phase transition and the creation of spontaneous curvature. These results highlight a general strategy to create stimuli-responsive polymersomes based on the fabrication of asymmetric membranes. For such membranes, polymersome bursting is driven by a change in membrane spontaneous curvature instead of an increase in membrane tension or a chemical degradation of block copolymers. UV light was successfully used as the stimulus to trigger the polymersome bursting. Nevertheless, temperature or electric or magnetic fields could also act as remote stimuli provided that one of the two leaflets of the membrane is composed of suitably designed LC copolymers. Although the thermo-responsiveness of LC polymersomes discussed was not driven by the LC transitions because of the dominance of the temperature-dependent solubility of PEG block, we believe if a suitable non thermo-sensitive hydrophilic block is used the general strategy to create stimuli-responsive polymersomes based on asymmetric membranes can also be implemented by thermo-stimulus. This design flexibility, combined with the low permeability of polymer bilayers, ensures a wide range of potential applications of LC polymersomes in the fields of drug delivery, cosmetics, and material chemistry.

**Acknowledgements.** We thank Jing YANG (BUCT, Beijing China) for her contribution to the polymer synthesis. We thank Daniel Lévy and Aurélie Di Cicco (Institut Curie, Paris) for their help in TEM imaging. We thank Prof. Marianne Imperor (LPS, Orsay) for her help in SAXS at SOLEIL. This work received support from the French "Agence Nationale de la Recherche" (ANR-08-BLAN-0209) and the "Foundation Pierre-Gilles de Gennes pour la Recherche".